\documentstyle[11pt,newpasp,twoside,epsf]{article}
\def\edcomment#1{\iffalse\marginpar{\raggedright\sl#1\/}\else\relax\fi}
\reversemarginpar

\begin{document}
\title{The Extended R\,CrA Young Association \altaffilmark {1}} 
\altaffiltext {1}{Based on observations made under the Observat\'{o}rio Nacional-ESO agreement for the joint operation of the 1.52\,m ESO telescope and at the  Observat\'{o}rio do Pico dos Dias, operated by MCT/Laborat\'{o}rio Nacional de Astrof\'{\i}sica,  Brazil} 
\author{G.R. Quast, C. A. O. Torres} 
\affil{Laborat\'{o}rio Nacional de Astrof\'{\i}sica/MCT, 37504-364 
Itajub\'{a}, Brazil}
\author{R. de la Reza,  L. da Silva and N. Drake}
\affil{Observat\'{o}rio Nacional/MCT, 20921-030 Rio de Janeiro, Brazil}

%\email{germano@lna.br, beto@lna.br, ldasilva@eso.org, delareza@on.br, drake@on.br}

\begin{abstract}
Observing ROSAT sources in an area covering $\sim$30\% 
of the Southern Hemisphere, including the R\,CrA Association, 
we found evidences that this nearby association may be much larger
than previously thought. 
Although in the survey we found many young stars near the R\,CrA, 
there are 20 young stars with properties that characterize them 
as possibly belonging to the R\,CrA association. 
From the nine known members with measured proper motions and 
radial velocities, we obtain the mean space velocity components 
for the Association  relative to the Sun: 
\mbox{(U, V, W)\,=\,($-3.8 \pm 1.2, -14.3 \pm 1.7, -8.3 \pm 2.0$})\,km/s. 
The new young stars with similar space velocities are in a projected
diameter of $\sim$35\deg.  
At a distance of $\sim$100\,pc, this represents a size of $\sim$60\,pc, 
similar to the spread in the kinematical distances obtained 
assuming the above space velocity components. 
If the original velocity dispersion  during star formation was 
equal to the dispersion of the velocity vector moduli 
($\sim$3\,km/s), then the age of the association
should be $\sim$10\,Myr to reach this size.  
In this way, the classical T Tauri spectroscopic
binary V4046 Sgr could be a member of the association.
\end{abstract}

\section{Introduction}
In Torres et al. and de la Reza et al. in these proceedings we describe
the SACY project. 
There we present the GAYA, a large  nearby young 
association near the South Pole. 
Another of the possible associations examined was that around the exotic
spectroscopic binary (SB) classical T\,Tauri star (CTT) V4046\,Sgr. 
We noted then that around the R\,CrA association (CAA) 
there are many young stars with similar kinematics forming a 
complex about 60\,pc in diameter. 
A similar work, deeper in X-ray emission but less in angular extent, 
has been presented by Neuh\"{a}user et al. (2000).

Frinck (1999) had already suggested that some of the stars associated 
with ROSAT sources could be PMS stars ejected from star forming clouds.  
The population of X-ray sources around star forming clouds seems to be represented by a mixture of PMS and young MS objects.
 
\section{Results}

\begin{figure}
\plottwo{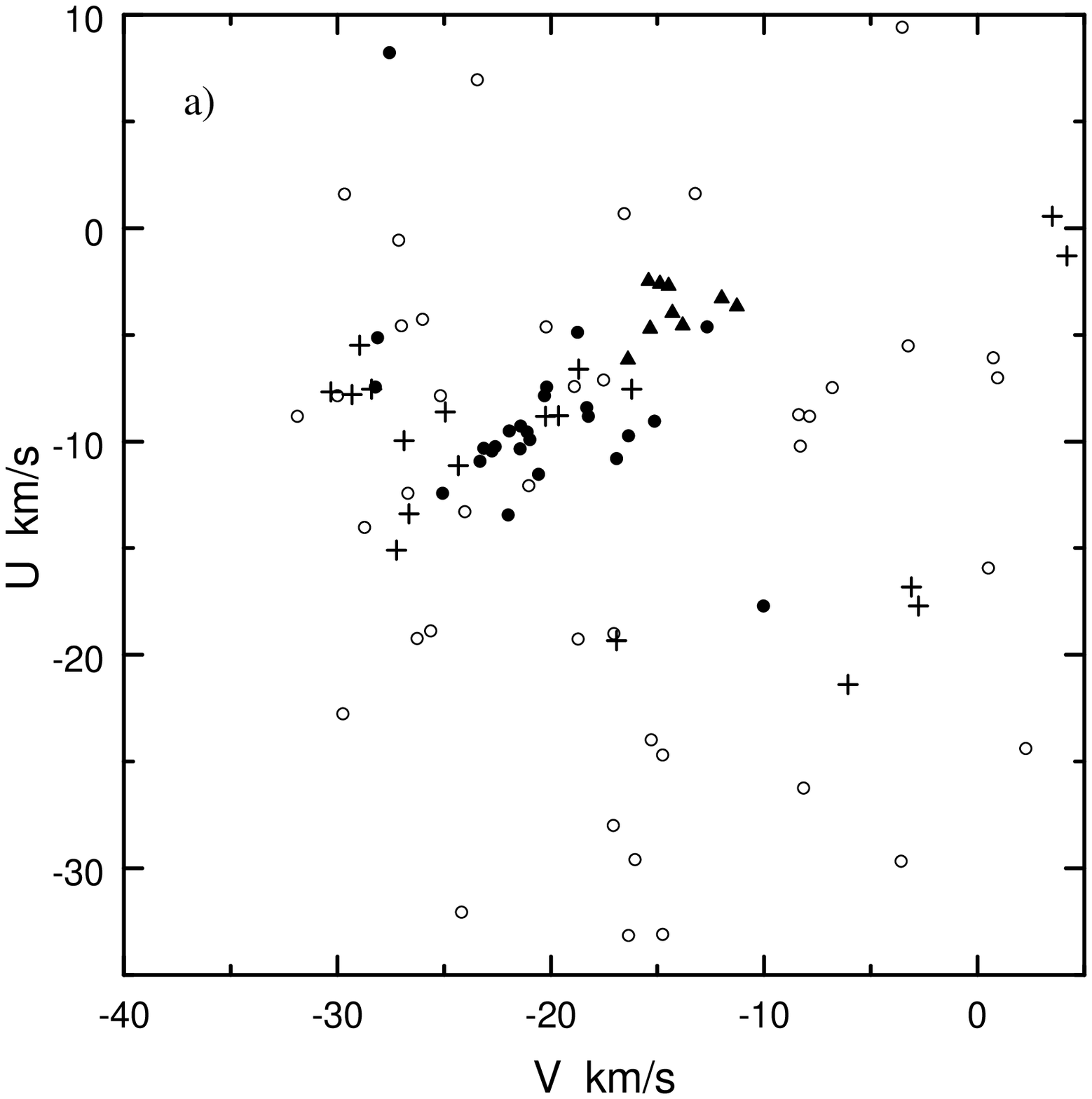}{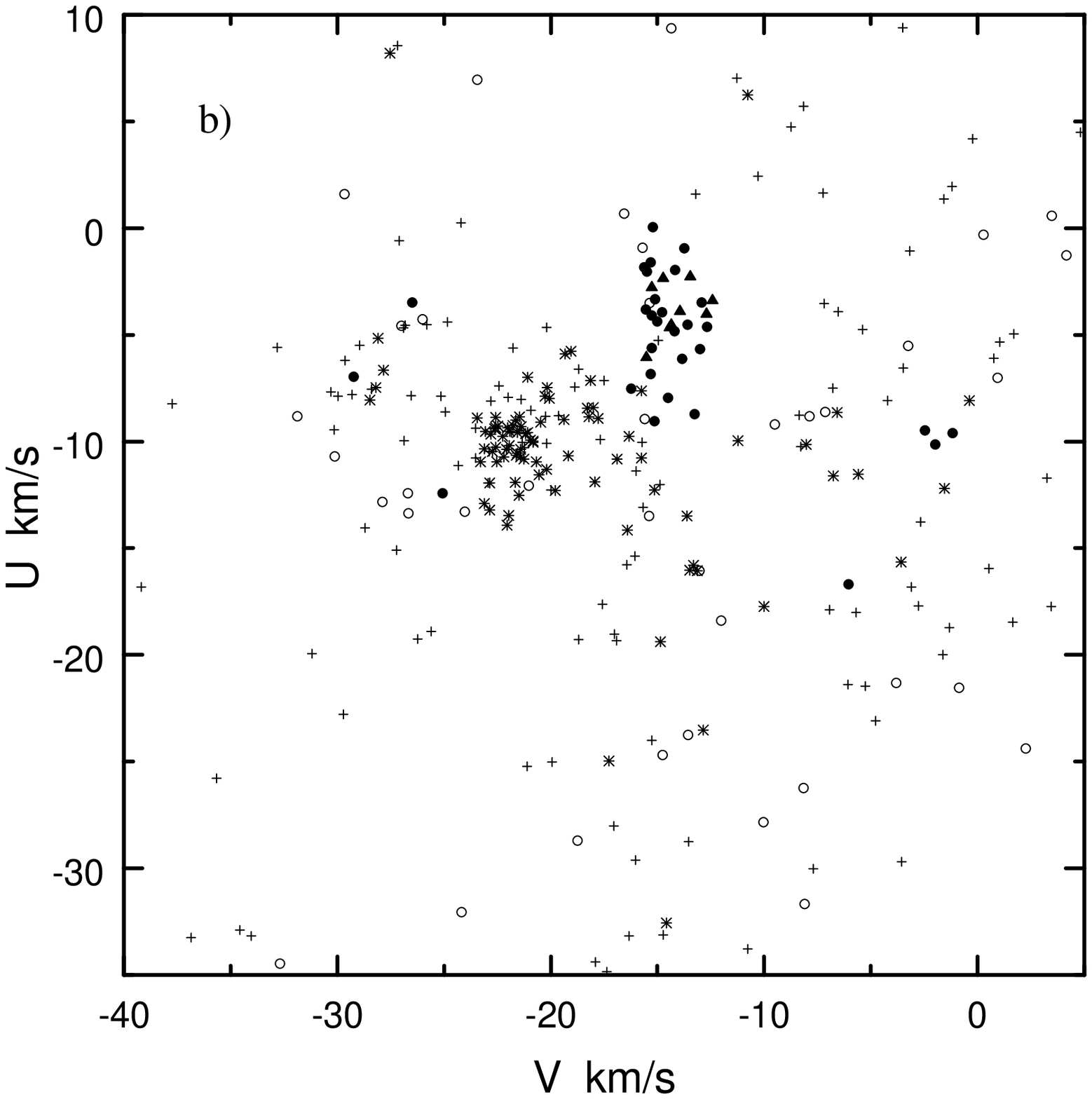}
\caption{UV space velocities for the stars observed in the SACY. 
a) Hipparcos stars - 
filled circles, plus and open circles represent stars with 
Li lines stronger, similar or weaker (or absent) 
than the Pleiades, respectively; 
triangles, bona fide R\,CrA association members.\\
b) All stars, including those with kinematical parallaxes. 
Stars within 17H\,$<$\,RA\,$<$\,21H have parallaxes estimated 
with the convergence point of the R\,CrA association - 
full circles  for strong Li, open for the others.
For the other stars the convergence point is that of the GAYA - asterisks for
strong Li and plus for the others.}
\end{figure}

In addition to the stars in the SACY we observed also some hot members 
of the CAA to obtain their radial velocities. 
From these data and those in the literature we established a set of
nine bona fide CAA members with known radial velocities and proper motions 
(we re-analysed some cases for this work). 
Adopting the distance of the CAA as 129\,pc  (Marraco \& Rydgren, 1981) -
five bona fide members that have been measured by Hipparcos 
have parallaxes consistent with this value - 
we obtained the space velocities:\\
\centerline{(U, V, W)\,=\,($-3.8 \pm 1.2, -14.3 \pm 1.7, -8.3 \pm 2.0$)\,km/s}

\begin{figure}
\plotone{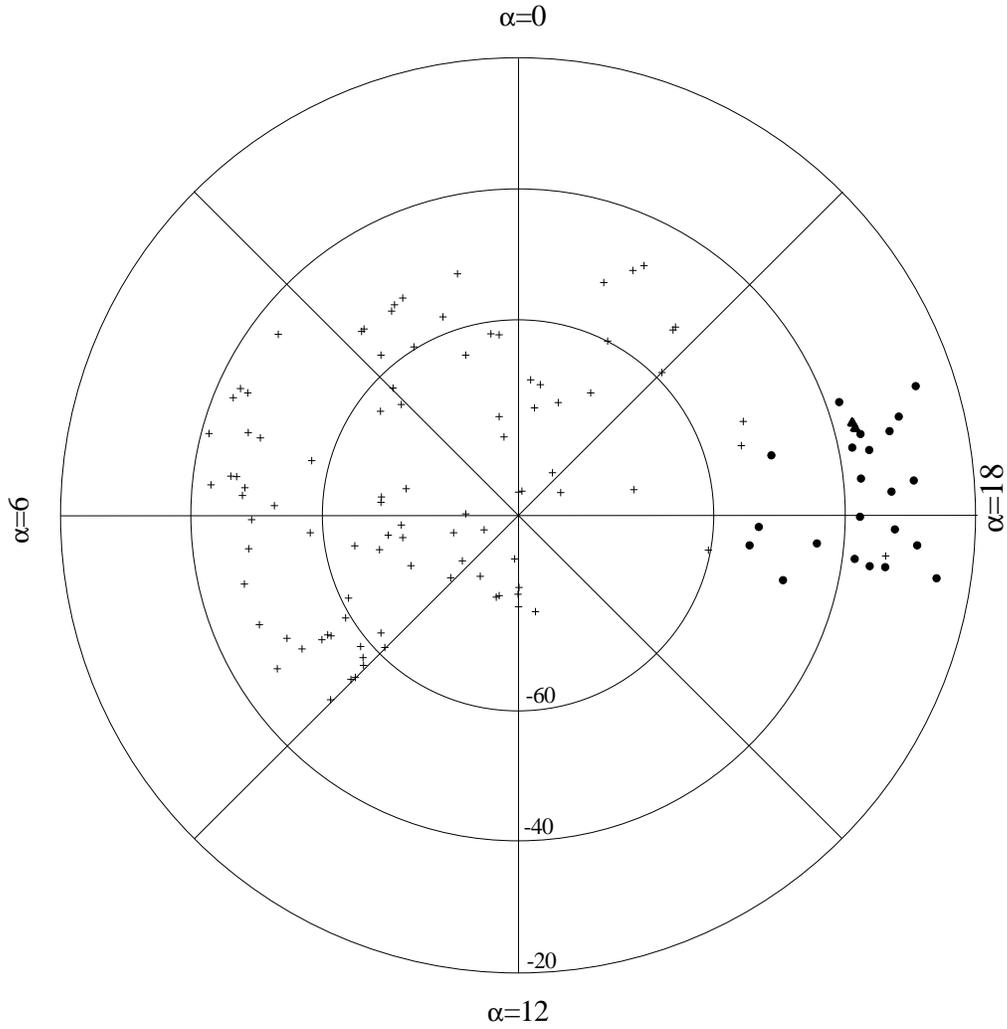}
\caption{Celestial polar projection of the young stars observed in SACY. 
Circles are probable members of the proposed 
extension of the R\,CrA association. 
Triangles are the bona fide members the R\,CrA association.  As it is very compact in this scale there is a great superposition of triangles.
Plus are for the other young stars (Li I lines stronger than the Pleiades).
Note that almost all young stars near the R\,CrA association are kinematically probable members of its extension. 
PZ Tel, proposed by Zuckerman and Webb (2000) to be a TucA member, is one of
the probable members of the extended association.}
\end{figure}

In Figure\,1a we show the (U,V) plane for these members and the SACY's stars
having trigonometric parallaxes. 
The position of the CAA in this plane is distinct of any other concentration
and there are very few other selected Hipparcos stars near this position. 
Although the above scattering  3\,km/s 
could be explained by  observational errors, 
it may also reflect a large 
original velocity dispersion during star formation.
We analysed the observations as in Torres et al. in these proceedings.
Taking the above space velocities, we used the kinematical method, 
described in Torres et al. (2000), 
to estimate the distances and space velocities for the stars with no 
reliable Hipparcos parallaxes.  
The concentration in the CAA position in the (U,V) 
plane is enhanced (Figure\,1b). 
As only stars within 17H\,$<$\,RA\,$<$\,21H are in this position (Figure\,2) - 
exactly where there are no proposed members for the GAYA -
for sake of visualization in Figure\,1b we converge the
stars outside the above range of RA to the UVW of the GAYA.  
The concentration is mainly formed by the nine bona fide stars and  
21 stars from SACY (four in common to Neuh\"{a}user et al. 2000).
In Figure\,2 it is clear that these 21 new probable members may be
very far from the clouds of the CAA, forming an ``extended CAA'' (ECA).

It is yet difficult to be sure about the boundaries of the ECA, 
mainly to the west where our survey began.  
In fact, the extension is mainly in that direction and it should be
noted that this is the correct one for the ``external agent'' 
that must have influenced the appearance of 
the CAA cloud complex (Graham, 1991).  
The very interesting SB2 CTT V4046\,Sgr (not a X-ray source)
may be a member of the ECA.  
In that case we obtain a kinematical 
parallax of 15.2\,mas,  somewhat larger than the value of 12.0\,mas
obtained from evolutionary scenarios (Quast et al. 2000).

The mean distance of the 21 new members is 100\,pc (Figure\,3a). 
Observational bias will favor the discovery of nearby members, 
but the spread in distance of  $\sim$80\,pc is roughly consistent 
with the apparent size of the ECA.  
In fact, the 30 stars are distributed in a projected area 
with a diameter of $\sim$35\deg, that is, a size of $\sim$60\,pc.

If the original velocity dispersion during star formation 
was similar to the average modulus
of the velocity vectors ($\sim$3\,km/s), 
it would take $\sim$10\,Myr to reach the size of $\sim$60\,pc.
This is similar to the evolutionary age (Neuh\"{a}user et al. 2000).
Actually, we deduced an age somewhat larger ($\sim$15\,Myr) as 
we are using closer distances. 
But these values are within the uncertainties. 
Anyway, the new members seem to be older than the bona fide ones.
On the other hand, Figure\,3a suggests that the ECA is expanding.

One of the best confirmations of the PTT nature of the proposed members 
is the behavior of the Li lines.  
In the same spectral range the previously known TTS in the CAA have a 
similar W$_{\it Li}$ distribution to the ECA showing that there is no 
significantly different Lithium depletion between them (Figure\,3b). 
We re-analysed the Li abundances of some of these TTS. 
Like other authors we obtained  abundances from the Li resonance 
line $\lambda$6707 above the interstellar ones. 
But using the high quality FEROS spectra we could use the 
$\lambda$6103 line, which is very sensitive to abundances, 
and the obtained values dropped to the interstellar ones.

\begin{figure}
\plottwo{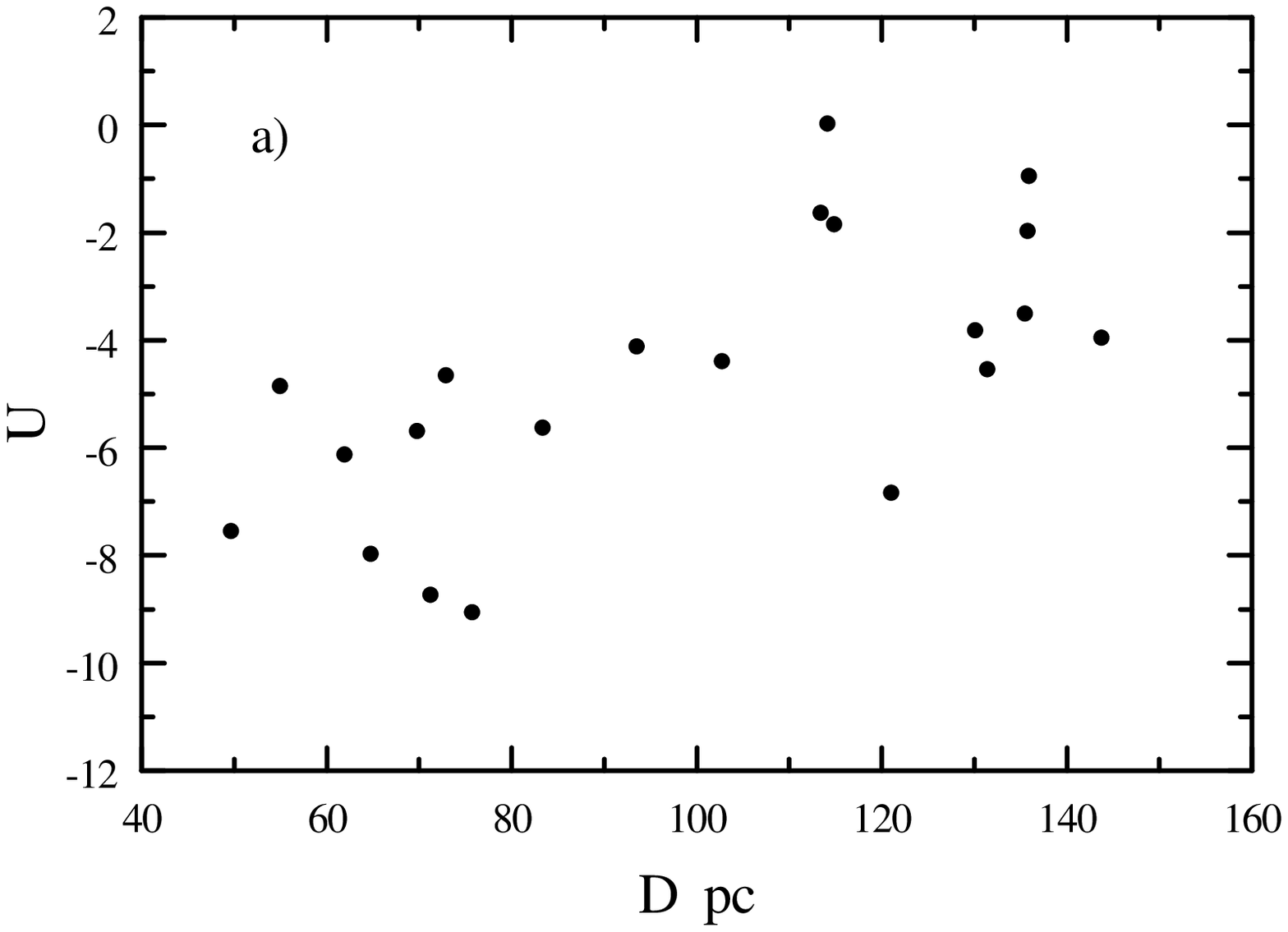}{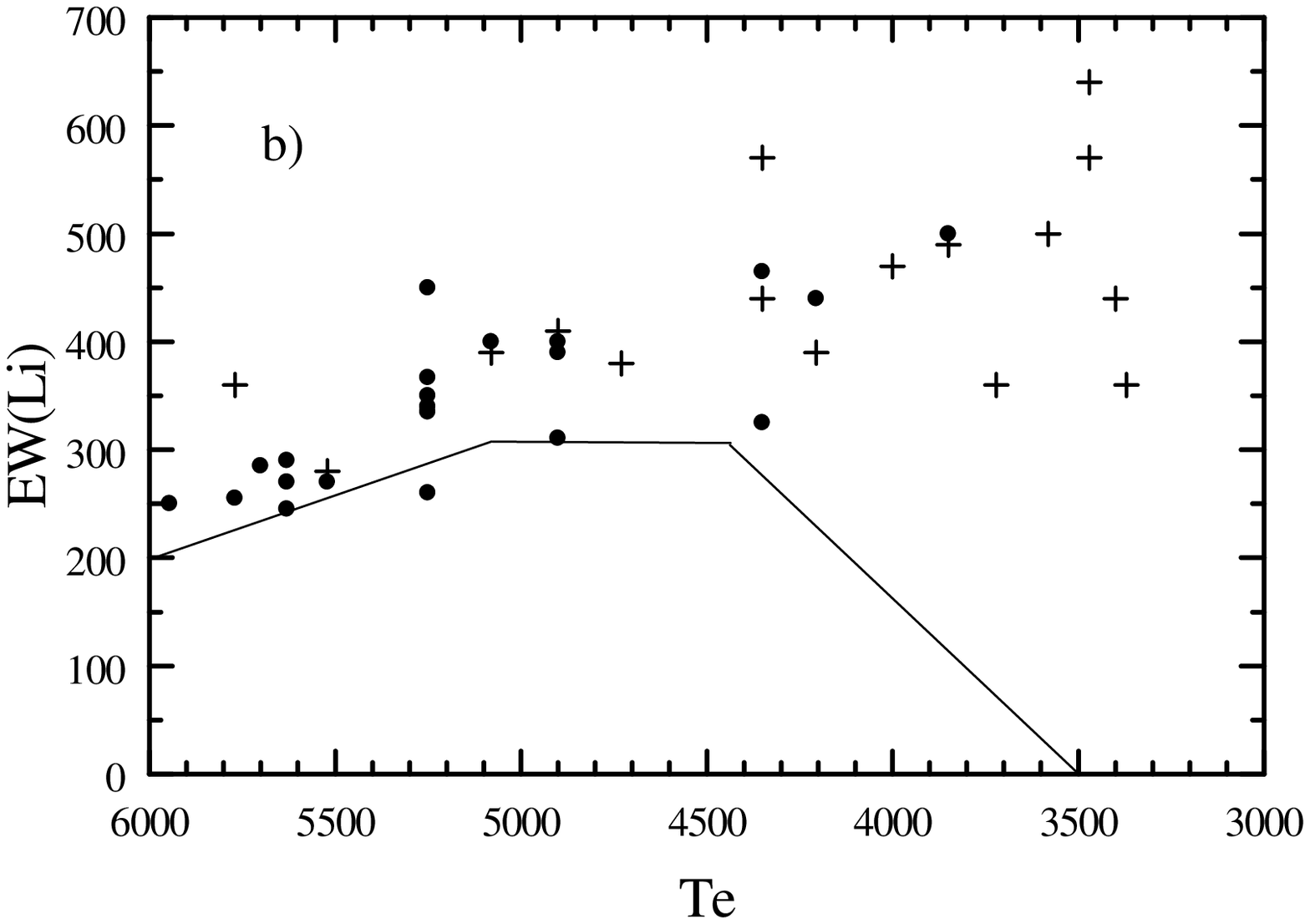}

\caption{a): U velocites as a function of the obtained distances for the probable members of the extended R\,CrA association. 
There is an indication that
this proposed association is in expansion.\\ 
b):   W$_{\it Li}$  (m\AA)  as a function of temperature for the
extended (circles) and the R\,CrA (plus) associations.}
\end{figure}

\section {Conclusions}

Exploring a region covering  about 30\% of the Southern Hemisphere 
we  found  evidences of a great extension of the young CAA 
and it seems to include the very interesting
SB2 CTT V4046 Sgr.  
The extended region is presently represented by at least 20 members,
having an age of $\sim$10\,Myr.

The distances of its members cover an interval from nearly 60 to about 
140\,pc (at a mean distance of $\sim$100\,pc) giving a diameter of
$\sim$80\,pc, compatible with the size produced after $\sim$10\,Myr
by an initial velocity dispersion of $\sim$3\,km/s.  
If real, the early CAA was a very turbulent environment.  
 
Almost all young stars in this region are kinematically possible members.  Although we found many other young stars in the vast
observed austral sky region, 
only around the CAA there are viable candidates for the ECA.  
This seems very hard to be a chance coincidence and we propose that the 
CAA complex is a very large and young
association, resulting from a turbulent initial phase, 
that is now visualised by the cloud format.
The ECA may be connected with the Upper Scorpius association, 
as both  have similar distances and ages, forming a spatial sequence
(see Sartori, in these proceedings).

\acknowledgments

C. A. Torres thanks FAPEMIG, G. R. Quast CNPq and R. de la Reza CAPES  for providing financial support.
This work was partially supported by a CNPq grant to L. da Silva 
(pr. 200580/97).
We thank M. Sterzik and M. Mayor for instructive discussions 
and C. Melo for some unpublished data.

\end{document}